# A redox-responsive hyaluronic acid-based hydrogel for chronic wound management

Ziyu Gao,[ab] Ben Golland,[b] Giuseppe Tronci [*bc] and Paul D. Thornton [**a]

[a] School of Chemistry, University of Leeds, Leeds, LS2 9JT, UK. ** Email address: p.d.thornton@leeds.ac.uk

[b] Biomaterials and Tissue Engineering Research Group, School of Dentistry, St. James's University Hospital, University of Leeds, UK. * Email address: g.tronci@leeds.ac.uk

[c] Clothworkers' Centre for Textile Materials Innovation for Healthcare, School of Design, University of Leeds, UK

## Abstract

Polymer-based hydrogels have been widely applied for chronic wound therapeutics, due to their well-acclaimed wound exudate management capability. At the same time, there is still an unmet clinical need for simple wound diagnostic tools to assist clinical decision-making at the point of care and deliver on the vision of patient-personalised wound management. To explore this challenge, we present a one-step synthetic strategy to realise a redox-responsive, hyaluronic acid (HA)-based hydrogel that is sensitive to wound environment-related variations in glutathione (GSH) concentration. By selecting aminoethyl disulfide (AED) as a GSH-sensitive crosslinker and considering GSH concentration variations in active and non-self-healing wounds, we investigated the impact of GSH-induced AED cleavage on hydrogel dimensions, aiming to build GSH-size relationships for potential point-of-care wound diagnosis. The hydrogel was also found to be non-cytotoxic and aided L929 fibroblast growth and proliferation over seven days in vitro. Such a material offers a very low-cost tool for the visual detection of a target analyte that varies dependent on the status of the cells and tissues (wound detection) and may be further exploited as an implant for fibroblast growth and tissue regeneration (wound repair).

**Introduction**

Polymeric hydrogels have emerged as highly promising materials for a range of biomedical applications, including carriers for controlled drug delivery,[1-3] protein adsorption,[4] the creation of contact lenses,[5,6] and as injectable implants.[7,8] The water-dominated composition of hydrogels renders them suitable for deployment in vivo, ensuring that such non-cytotoxic materials are also excellent candidates for use as scaffolds that facilitate tissue regeneration.[9-12] In addition, hydrogels that undergo an actuated change in dimensions in response to a target analyte may act as biosensors. Such materials have been deployed for the detection of bacterial enzymes,[13] toxins,[14] disease specific genes,[15] and pathogenic proteins that are associated with particular disease states.[16] Polymer hydrogels therefore hold great promise as materials for chronic wound management, both for the detection of chronic wound-associated markers as a mode of diagnosis, and by providing a scaffold for fibroblast growth, and subsequent new tissue generation, as part of wound healing.

Chronic states, such as those associated with non-self-healing ulcers, exhibit excessive oxidant stress, leading to upregulated reactive oxygen species (ROSs), inhibited cell migration and proliferation, and growth factor deficiency.[17] Whilst antioxidants regulate endogenous oxidant production in active wounds so that repair occurs via an orderly process, nonself-healing ulcers are in a persistent inflammation state, whereby prolonged infiltration of polymorphonuclear leukocytes and mononuclear cells results in reduced concentration of antioxidants, for instance glutathione (GSH), and incorrect redox potential in the cells.18 Tracking biochemical shifts, such as antioxidant concentration, could therefore be useful to diagnose changes in wound chronicity and assist with clinical decision making. Given that wound diagnosis relies on clinical assessment,[19] hydrogels with inherent oxidation responsivity could play a significant role as a cost-effective point-of-care diagnostic system. Following hydrogel contact with the wound, oxidation-induced changes in hydrogel size could be exploited to describe defined chronic wound states and inform clinical therapies.

Hyaluronic acid (HA), a natural polysaccharide that is a major constituent of the extracellular matrix, offers biocompatibility, anti-adhesivity, biodegradability, and non-immunogenicity.[20] HA content within the human body is approximately 15 g (for a 70 kg human), and largely resides in the skin and musculoskeletal tissue.[21] The hydrophilicity of HA ensures that it is an excellent candidate for use as a gelator within both physical[22] and chemical hydrogels.[23] HA-containing hydrogels have been reported as effective scaffolds for the regeneration of bone,[24] cartilage,[25] neural tissue,[26] and myocardial tissue.[27] In direct relation to wound healing, recent examples of HA-based chemical hydrogels include an anti-oxidative and anti-inflammatory drug release system for early tendinopathy intervention,[28] an in situ-forming biomimetic dressing for soluble factor-free wound healing,[29] a UV-cured drug-encapsulated antibacterial hemostat,[30] and injectable hydrogels for the management of osteoarthritic synovial fluids.[31] However, despite substantial efforts being afforded to the employment of HA as a component of a therapeutic system, limited reports have proposed the application of HA for chronic wound diagnostics, whereby expensive and sophisticated imaging apparatus are required.[32] GSH, a natural tripeptide that consists of glutamate, cysteine and glycine, is abundant in the majority of animal cells. GSH plays a key role in biological processes including biocatalysis, metabolism, signal transduction, gene expression, apoptosis and anti-oxidation.[33] GSH is present in both its reduced (GSH) and oxidized state (GSSH), and the ratio of the two variants dictate many pharmacological processes within the cell. The thiol that the cysteine unit presents ensures that GSH is a reducing agent and offers a signal for GSH detection.[34] The intracellular concentration of GSH is typically 0.5 mM for normal cells, compared to 10 mM for cancerous cells.[35] On the other hand, averaged GSH levels were found to be lower in diabetic foot ulcers (0.0534 nmol·mg$^{-1}$ wet tissue weight) in line with the excessive oxidant stress in the former case.[36] Consequently, materials that are able to respond to GSH concentrations within this range hold great promise for the detection of pathological GSH concentrations that may be linked to either the onset of cancer or wound chronicity.

In this paper we describe the facile, one-step synthesis of a HA-based chemical hydrogel that is spatially responsive to the presence of GSH. Aminoethyl disulfide (AED) is employed as a dynamic covalent crosslinker that is sensitive to the presence of GSH; AED-GSH conjugation results in crosslinker cleavage, increased hydrogel swelling and a clearly visible change in hydrogel dimensions. The hydrogel was found to be noncytotoxic and aided fibroblast growth and proliferation over seven days. Such a material offers a very low-cost method for the visual detection of a target analyte that varies dependent on the status of the cells and tissues (wound detection), and may be further exploited as a scaffold for fibroblast growth and tissue regeneration (wound repair).

**Experimental**

**Materials and methods**

Hyaluronic acid sodium salt (molecular weight: 1200 kDa, cosmetic grade) was purchased from Hollyberry Cosmetic, 4-(4,6-dimethoxy-1,3,5-triazin-2-yl)-4-methyl-morpholinium chloride (DMTMM) and 2-(N-morpholino)ethanesulfonic acid (MES) were purchased from Fluorochem. AED, rhodamine B isothiocyanate-dextran (RITC-dextran) and ninhydrin reagent were purchased from Alfa Aesar. Alamar Blue assay kit was purchased from ThermoFisher Scientific. All other reagents were purchased from Sigma-Aldrich.

**Preparation of hyaluronic acid hydrogels**

HA powder was dissolved in deionised water to prepare a 2 wt% HA solution. MES buffer (0.1 M) was added at room temperature under stirring to provide a weak acid environment (pH 5.5). DMTMM (2 equivalents per HA repeat unit) was then added at 37 °C to activate the carboxyl groups of HA. The temperature was maintained at 37 °C for 1 hour, and a molar ratio of either 0.2 or 0.4 moles of AED relative to the moles of each HA repeat unit was added to the solution (Table 1). The stirring speed was increased to 1000 rpm for 5 minutes and either 0.6 g or 0.8 g of the reacting solution was cast into either 12- or 24-well plates, respectively. HA hydrogels were obtained after 2-hour incubation at 37 °C.

Table 1. The quantities of AED used to make two covalent hydrogels

| Sample ID | HA (wt%) | $[AED][COOH]^{-1}$ |
|---|---|---|
| 1 | 2 | 0.2 |
| 2 | 2 | 0.4 |

**Analysis of HA crosslinking**

Attenuated total reflectance Fourier transform infrared (ATRFTIR) spectroscopy was used to characterise the crosslinked structure of HA hydrogels. IR spectra were obtained from freeze-dried networks and recorded between 4000 and 500 cm$^{-1}$ using a Bruker spectrophotometer at room temperature. The thermal properties of both HA and the freeze-dried networks produced were evaluated by differential scanning calorimetry (DSC) using a DSC Q20 unit (TA instruments) calibrated with indium. DSC thermograms were recorded under nitrogen atmosphere with a heating rate of 10 °C·min$^{-1}$ from 30 °C to 200 °C.

The progress of the crosslinking reaction was investigated by 2,4,6-trinitrobenzenesulfonic acid (TNBS) and ninhydrin assays.[37,38] TNBS was used to quantify any unreacted AED entrapped in the crosslinked network. 0.8 g of newly synthesised hydrogel was freeze-dried and immersed in 2 mL NaHCO$_3$ (4 wt%) at 40 °C for 30 minutes to wash away any unreacted AED; 1.0 mL of the supernatant was collected and investigated by TNBS assay. 1.0 mL TNBS solution (0.5 wt%) was added to the supernatant in the dark and incubated at 40 °C for 3 hours with a rotation speed of 120 rpm. 3 mL HCl (6 N) was added to the incubated solution and temperature raised to 60 °C for 1 hour to terminate the reaction. After solution equilibration to room temperature, the sample solutions were diluted with 5 mL H$_2$O. The unreacted TNBS was washed out by extraction with 20 mL diethyl ether, three times. 5 mL of the retrieved sample solution was immersed in hot water to evaporate any diethyl ether, and diluted with 15 mL H$_2$O. Finally, 2 mL of solution was analysed by UV-Vis spectroscopy at 346 nm. Quantification of any AED residue was carried out by comparison with an AED calibration curve.

The presence of any terminal free amino group deriving from the grafting, rather than the complete crosslinking, of HA with AED was quantified by ninhydrin assay. 10 mg of washed, freeze-dried polymer was immersed in 1 mL water. 1 mL ninhydrin solution (8 wt%) was then added in the dark, and the solution quickly immersed in a water bath at 95 °C for 15 minutes. 1 mL ethanol was added to stop the reaction after equilibration at room temperature. A calibration curve was obtained by measuring UV-Vis absorbance at 348 nm using various amounts of glycine.

**Hydrogel swelling and degradation studies**

Newly-synthesised hydrogels were weighed ($m_w$) and then immersed in deionised water. As control samples, another series of hydrogels was soaked in H$_2$O with 5 mM GSH to determine any hydrogel response. The mass of the samples ($m_t$) was recorded at selected time points (1, 2, 3, 4, 7, 14, 21, 28 days) following light sample blotting with tissue paper. The swelling ratio (SR) of each hydrogel was calculated using the
following equation:

$$SR = \frac{m_t}{m_w} \times 100$$

Hydrogel degradability was evaluated via gravimetric analysis of covalent networks following incubation in aqueous solutions and drying. The initial network mass of hydrogel 1 was recorded as $m_0$. 0.8 g of freeze-dried hydrogel 1 freeze-dried sample was immersed in PBS solution either with GSH (5 mM) or without GSH. Samples were collected following 1, 2, 3, 4, 7, 14, 21 and 28 days, washed with deionised water, freeze-dried and weighed ($m_d$). The percentage of the crosslinked network remaining ($μ_{rel}$) was calculated via the following equation:

$$\mu_{rel} = \frac{m_d}{m_0} \times 100$$

**Mechanical tests**

Hydrogels were tested using a rheometer (Anton Paar, MCR 302) with a 25 mm parallel plate. Frequency sweeps were recorded at 100–1 rad·s$^{-1}$ with a constant amplitude (1%) at room temperature, using a 0.8 g hydrogel and a 2 mm gap. Time sweep measurements were carried out to quantify hydrogel gelation kinetics following the reaction of HA with AED. 0.6 mL of solution was injected homogenously on the sample holder and tested with a 25 mm parallel plate. The measurement was carried out at 37 ˚C over 150 minutes with constant amplitude (1%) and frequency (5 Hz).

Compression measurements were conducted using Bose ELF 3200 apparatus. All the hydrogel samples (Ø: 3 mm; h: 3 mm) were tested with a compression speed of 0.02 mm·s$^{-1}$.

**Scanning electron microscopy (SEM)**

The network microstructure of newly synthesised and partially degraded networks was inspected by SEM using a JEOL JSM6610LV microscope under 5 kV voltage following gold coating. Freeze dried samples were collected following either synthesis or 2-day immersion in 20 mL solution of either PBS or GSH (5 mM)-supplemented PBS. Pores (n = 20–60) were visualised and their diameter measured.

**Macromolecule uptake studies**

Hydrogels were loaded with RITC-dextran, (Mn: 70 000 g mol_1), as a model macromolecule. 0.6 g hydrogel was immersed in 2 mL RITC-dextran-containing (0.5

mg·mL$^{-1}$) PBS solution at 37 1C for 24 hours. Samples were washed with PBS solution to remove any surface-coated RITC-dextran before being transferred into 3 mL PBS solution that contained various GSH concentrations (0–20 mM). RITC-dextran uptake was calculated through the difference in supernatant absorbance before and after RITC-dextran loading (calibration equation: y = 0.0011x – 0.004, $R^2$ = 0.9995, y = absorbance, x = RITC-dextran concentration (µg·ml$^{-1}$)).

The change of hydrogel structure, and the distribution of RITC-dextran loaded in the hydrogel, was investigated by analysis of the lyophilised networks of hydrogels that had undergone incubation in PBS solution that contained or lacked SH at 37 °C, by laser scanning confocal microscopy (LEICA TCS SP8, excitation wavelength 552 nm).

**Hydrogel response to simulated wound fluid**

Simulated wound fluid was prepared as reported.[39] Briefly, 5.844 g sodium chloride (NaCl), 3.360 g sodium hydrogen carbonate (NaHCO3), 0.298 g potassium chloride (KCl), 0.2775 g calcium chloride (CaCl2) and 33.00 g bovine serum albumin (BSA) were dissolved into 1 L of deionized water. 0.6 g of hydrogel was immersed in 20 mL SWF loaded with varied GSH content (0-20 mM) to mimic multiple wound chronic states. All the samples were incubated at 37 °C, stirred at 100 rpm and any change in hydrogel size recorded following 24-hour incubation.

**Cytotoxicity evaluation**

Hydrogel (8 mm$^3$) was incubated in 70 wt% ethanol (×3), and then washed in PBS buffer (×3), and ultimately in cell culture medium (×3). L929 murine fibroblasts were cultured (37 1C, 5% $CO_2$) in Dulbecco's modified Eagle's medium (DMEM) supplemented with 10% fetal bovine serum (FBS), 1% glutamine, and 0.5% penicillin-streptomycin. The cell suspension ($1\times10^4$ cells·mL$^{-1}$) was transferred to a 96-well-plate (200 mL·well$^{-1}$), followed by application of individual hydrogel samples to each well. Cell viability was quantified by Alamar Blue assay after 1-7–day culture. Cells cultured on either tissue culture plates (TCPs) or non-treated tissue culture plates (NTCPs) were set as a positive control and a negative control, respectively. Four replicates from four individual hydrogels were analysed.

**Statistical analysis**

For statistical analysis, all the cytotoxicity results were analysed at least three times. The significant difference was calculated through One ANOVA analysis with a p level at 0.05. The results are presented as *$p < 0.05$, **$p < 0.01$, ***$p < 0.001$, ****$p < 0.0001$.

**Results and discussion**

Chemical hydrogels were successfully generated from a 2 wt% HA aqueous solution by crosslinking HA with either 20 mol% (hydrogel 1) or 40 mol% (hydrogel 2) AED, with respect to HA carboxylic acid groups (Scheme 1).

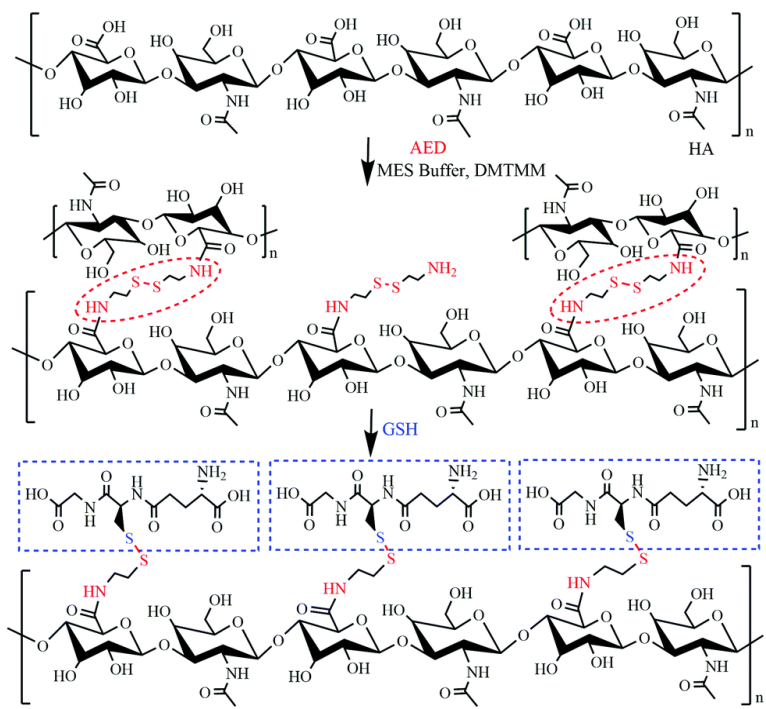

**Scheme 1.** The reaction between AED and HA chains yields a crosslinked polymer that forms a chemical hydrogel when maintained in aqueous solution. GSH-induced AED disulfide reduction disrupts the crosslinked network, resulting in changes to the hydrogel dimensions and mechanical properties.

As a result, HA chains are crosslinked via disulfide bridges, which equip the obtained hydrogels with wound diagnostic functionality. Contact of the hydrogel with GSH-containing wound exudate leads to the cleavage of disulfide bridges and covalent network disintegration, so that defined [GSH]:[hydrogel size] relationships can be built for wound diagnosis.

Qualitative confirmation of HA crosslinking was achieved by FTIR spectroscopy (Fig. 1); the spectrum of the crosslinked networks revealed a peak at 1700 cm$^{-1}$, which corresponds to the amide bonds formed between HA and AED. In addition, the peak present at 538 cm$^{-1}$ in the crosslinked samples is attributed to the presence of disulfide bonds that are present within AED molecules.

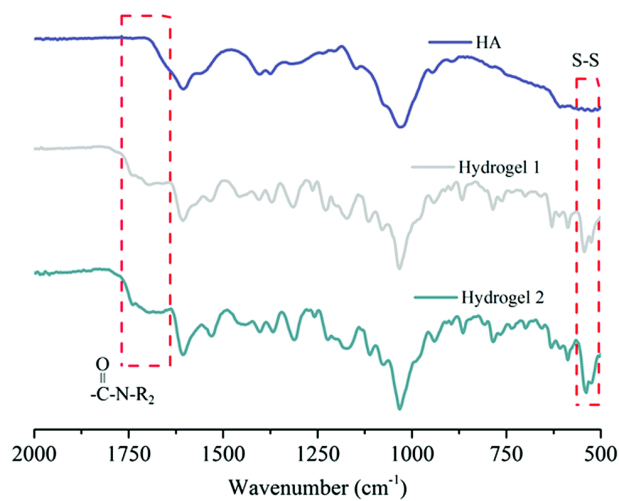

**Figure 1.** FTIR spectrum of HA (top), freeze-dried hydrogel 1 (middle) and freeze-dried hydrogel 2 (bottom). The peaks that confirm successful HA crosslinking are highlighted.

For hydrogel 1, TNBS and ninhydrin assays revealed that 82 mol% of AED amine groups had reacted to form a complete covalent crosslink, rather than forming AED-grafted residues. In the case of hydrogel 2, 61 mol% of amine groups presented by AED had reacted. DSC analysis revealed endothermic peaks at 138.9 °C (hydrogel 1) and 143.0 °C (hydrogel 2) in the thermograms of the crosslinked polymers that were absent from the thermogram corresponding to HA (Fig. S1, Supp. Inf.). These additional peaks may be ascribed to the thermal decomposition of the AED crosslinks.

Time-sweep rheometry was carried out to investigate the kinetics of the crosslinking reaction. Rheograms revealed that hydrogel 1 was formed with a storage modulus of 885 Pa and a loss modulus of 9 Pa after 130 minutes (Fig. 2), whilst hydrogel 2 was formed with a storage modulus of 1029 Pa and a loss modulus of 13 Pa after 140 minutes (Fig. S2, Supp. Inf.).

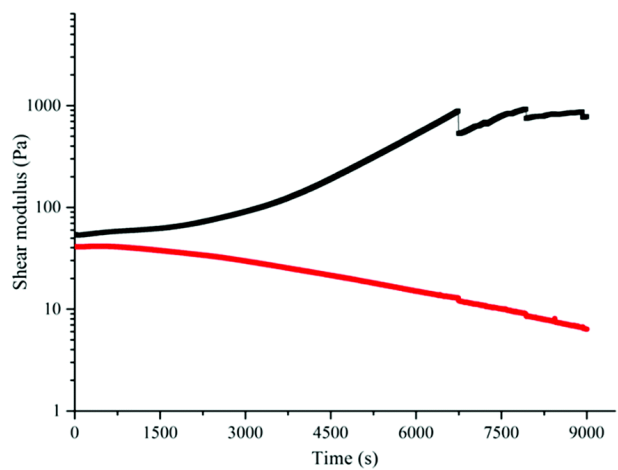

**Figure 2.** Time sweep rheological analysis during HA crosslinking to assess gelation kinetics for hydrogel 1. The upper, black, line corresponds to $G'$ and the lower, red, line corresponds to $G''$.

Such a result confirms that the hydrogels can be created in a simplistic manner within a relatively short timeframe; the increased gelation times displayed by hydrogels formed from reacting mixtures containing increased content of AED may be attributed to the occurrence of grafted rather than crosslinked HA chains, as supported by TNBS and ninhydrin results. These results give indirect evidence of the impact that the HA/AED molar ratio has on the network crosslinking yield. Overall, rheological analysis revealed that the storage modulus of both crosslinked materials exceeded the loss modulus, confirming their classification as hydrogels and the formation of a solid-like hydrogel rather than a viscous liquid (Fig. 3 and Fig. S3, Supp. Inf). Due to the superior crosslinking efficiency, hydrogel 1 was chosen for further investigations concerning its swelling capabilities, and its susceptibility to disassemble in the presence of the target analyte, GSH.

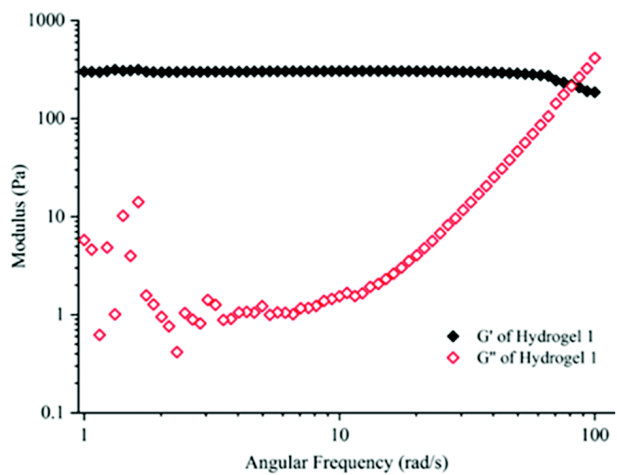

**Figure 3.** Frequency sweep rheological analysis revealing the storage (*G'*) and loss (*G''*) modulus of hydrogel 1.

When incubated in deionised (DI) water, newly-synthesised hydrogel 1 displayed a 700 wt% mass increase after 21 days (Fig. 4a). In contrast, when incubated in the presence of aqueous GSH solution (5 mM), hydrogel 1 started to break and could not be weighed after two days owing to the fragility of the material. This strongly advocates that GSH-induced cleavage of the disulfide bonds in AED-crosslinked HA chains occurs, resulting in the disintegration of the covalent network due to GSH-AED bond formation (Scheme 1). The mass of lyophilised polymer that had been maintained as a hydrogel in aqueous solution that contained GSH was greater than the mass of polymer that had previously been maintained as a hydrogel in aqueous solution that lacked GSH (Fig. 4b). The difference in the mass of recovered polymer is credited to GSH conjugation to the HA backbone. After four days storage in GSH solution the hydrogel could not be recovered, rendering lyophilisation and subsequent mass recording was impossible.

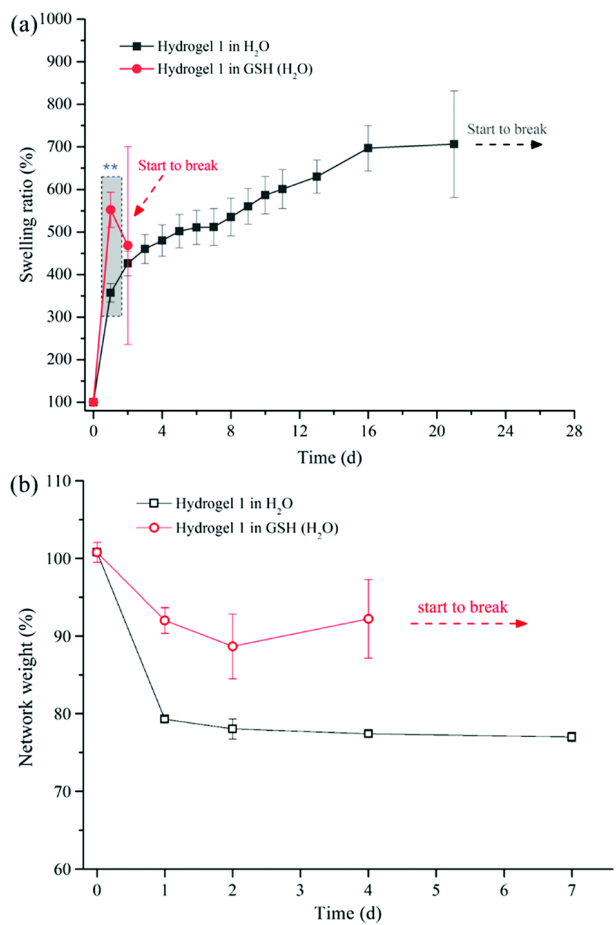

**Figure 4.** Hydrogel 1 swelling (a) and degradation (b) in DI water (black line) versus hydrogel 1 swelling in aqueous GSH solution (5 mM, red line).

Next, the compression properties of hydrogel 1 were investigated. Hydrogel 1 exhibited great compressibility as evidenced through stress–compression curves (Fig. 5).

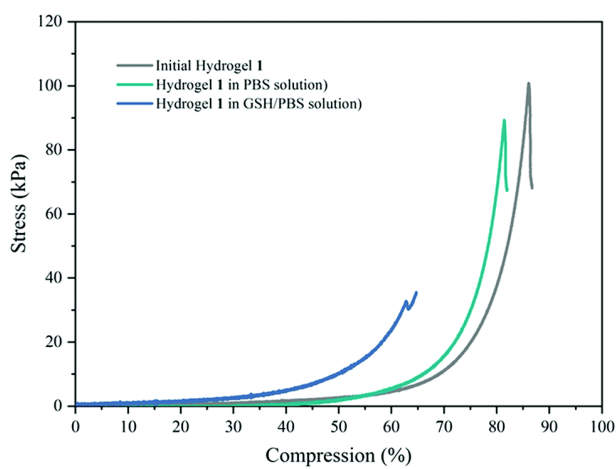

**Figure 5.** Stress–compression curves for hydrogel 1 following synthesis and 48-hour incubation in either PBS solution or GSH (5 mM)-supplemented PBS buffer solution.

Following 48-hour incubation in GSH (5 mM)-supplemented PBS solution, the stress at break of hydrogel 1 was significantly lower than the stress at break of both the newly-synthesised hydrogel 1, and hydrogel 1 following 48 hour incubation in GSH-free PBS solution (Fig. S4, Supp. Inf.). The original hydrogel could hold up to 82% compression with stress at break of 90 ± 8 kPa. After being immersed in PBS for 48 hours, a slight decrease in stress at break was observed ($\sigma_b$ = 75 ± 10 kPa) and the compression at break was also reduced to 80%. In the presence of GSH solution, a significant decrease in stress at break ($\sigma_b$ = 30 kPa), and compression at break ($\varepsilon_b$ = 62%) was observed after 48 hours, providing convincing evidence for GSH-mediated hydrogel disassembly.

The morphology of hydrogel 1 incubated in PBS or GSH- supplemented PBS were assessed by SEM (Fig. 6), to investigate whether GSH-induced network disintegration resulted in significant changes in material microstructure. SEM images were captured following 48-hour incubation, at which point the GSH-treated material was no longer stable in hydrogel form.

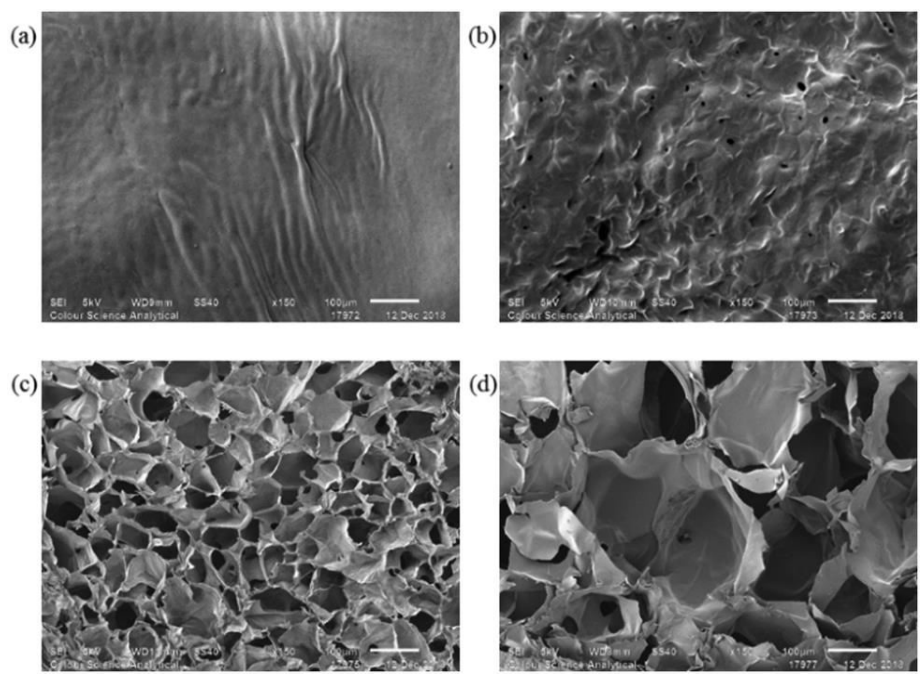

**Figure 6.** SEM images of hydrogel 1 captured following 48 hour incubation in either PBS solution (a and c), or GSH-supplemented PBS solution (5 mM) (b and d). Scale bar = 100 μm.

Clear differences in the morphologies of the two samples can be observed; in the absence of GSH, the material presents a predominantly smooth surface that largely lacks major defects, whereas in the presence of GSH the material contains substantial pores that are ascribed to GSH-mediated crosslinker cleavage (Fig. 6a and b, respectively). The internal structure of hydrogel 1 is porous, with an average pore diameter of 58 ± 18 µm (Fig. 6c). The internal pore diameter of the GSH-treated (48 hours) hydrogel was found to be 176 ± 89 µm, whereby the greater pore size again supports the rationale for the degradation effect that GSH has on the crosslinked polymer structure (Fig. 6d).

In order to apply the hydrogel within a (*ex vivo*) diagnostic setting, the uptake of RITC-dextran by hydrogel 1 was assessed by confocal microscopy. The hydrogels were incubated in RITC-dextran solution, prior to lyophilisation and analysis of the recovered polymeric network. When the hydrogels are incubated in greater concentrations of GSH, more extensive scaffold disruption, and increased gel porosity, arises due to the onset of network degradation. Consequently, the fluorescent marker is able to penetrate more readily in the partially-degraded network upon network cleavage induced by increasing GSH levels (Fig. 7).

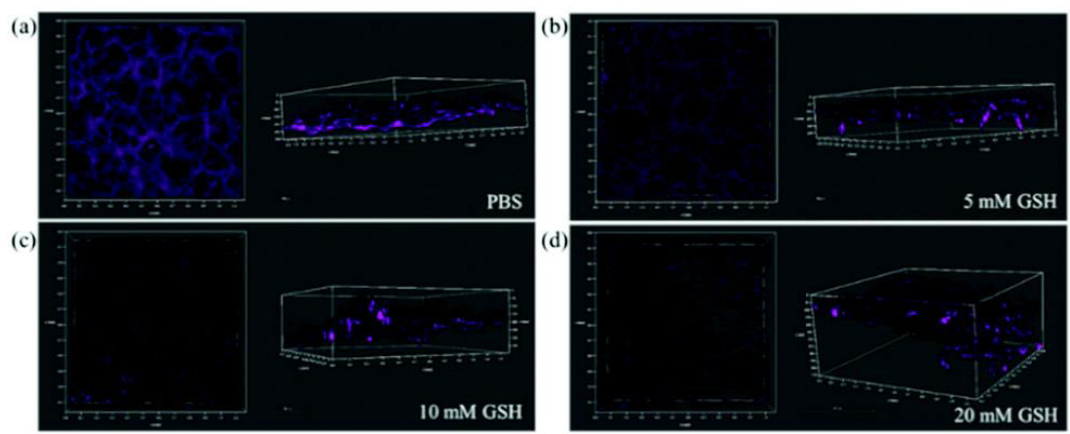

**Figure 7.** Confocal images hydrogel 1 loaded with RITC-dextran in (a) PBS solution, and PBS solution supplemented with (b) 5 mM GSH, (c) 10 mM GSH, and (d) 20 mM GSH. The stained polymer network is provided in the left-hand image of each box, and becomes increasingly disintegrated with increased GSH concentration. The depth to which RITC-dextran was able to penetrate the hydrogel was related fluorescence depth profile for each hydrogel is provided in the right-hand image of each box.

The depth to which RITC-dextran was able to penetrate the hydrogel was related to GSH concentration, increasing from 125 μm (No GSH) to 200 μm (5 mM), 250 μm (10 mM), up to 300 μm (20 mM GSH).

The pronounced effect that GSH has on hydrogel swelling may also be exploited to realise a diagnostic device that undergoes dimension changes that are visible to the naked eye upon interaction with GSH. Fig. 8 demonstrates the macroscopic changes in hydrogel 1 swelling that occur when the hydrogel is incubated in simulated wound fluids that contain various concentrations of GSH, as a mimetic of varying chronic wound states.

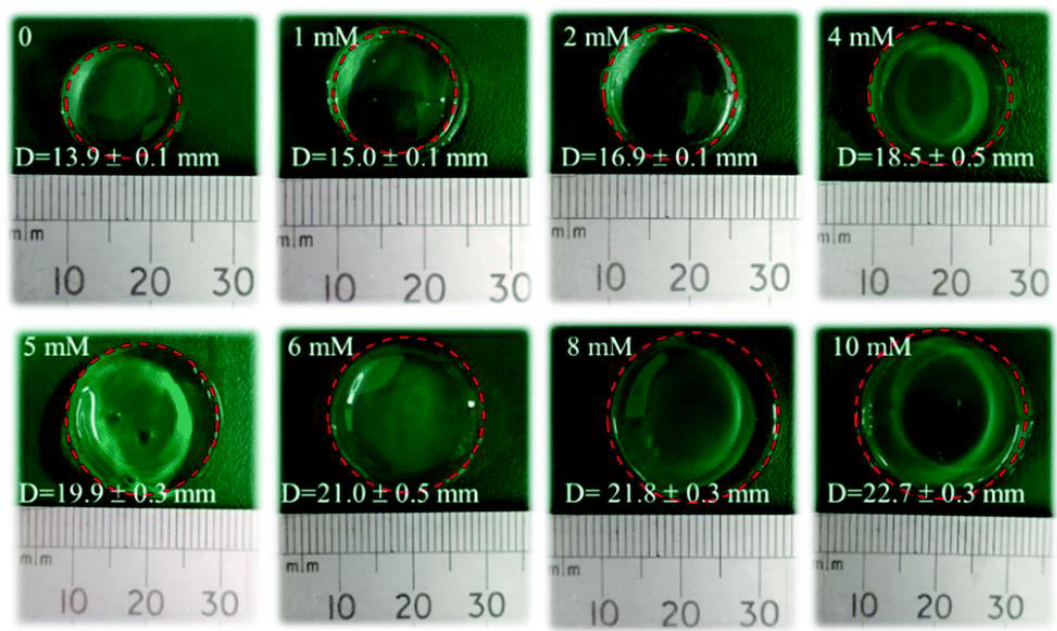

**Figure 8.** Hydrogel 1 swelling upon interaction with GSH. The extent of swelling is linked to the GSH concentration in the simulated wound fluid in which the hydrogel was stored. The edges of the three-dimensional hydrogels are circled for clarity.

As anticipated, hydrogel swelling increases with increasing GSH concentration due to a reduction in the covalent crosslinks that originally maintain the hydrogel. Increasing the GSH concentration from 0 mM to 5 mM results in an increase in hydrogel diameter by ~43%. Increasing the solution GSH concentration from 0 mM to 10 mM results in the hydrogel diameter increasing by 63%, and increasing the

GSH concentration from 0 mM to 20 mM results in hydrogel dissolution (Fig. S5, Supp. Inf.). As average GSH levels are found to be lower in diabetic foot ulcers (53.4 pmol·mg$^{-1}$ wet weight), compared to control tissues (150.6 pmol·mg$^{-1}$ wet weight), material expansion would be a sign of healthy tissue. Conversely, elevated GSH levels in tumour tissue compared to healthy tissue, and the accompanying increase in hydrogel dimensions, may act as a simple diagnostic test for the exposure of cancerous cells.[40]

To further probe the suitability of deploying hydrogel 1 as part of a wound management device, the cytotoxicity of the material was assessed using the Alamar Blue colorimetric assay for cell viability (Fig. 9).

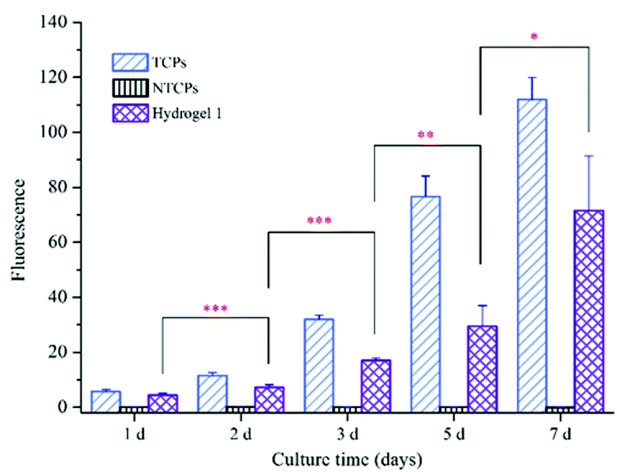

**Figure 9.** L929 cells viability after 7 day culture on TCPs, NTCPS and hydrogel 1.

It is essential that a diagnostic material is not detrimental to healthy cells in order to avoid false summations. Additionally, if the hydrogel is intended to be used *in situ* to aid wound healing whilst simultaneously providing diagnostic information, it is essential that the hydrogel promotes fibroblast growth and proliferation over an extended period. Cell viability is positively related to fluorescent intensity in the Alamar Blue assay. L929 cells grew most rapidly on TCPs, in contrast to no cells surviving following culturing on NTCPs. The cells presented excellent proliferation in the presence of hydrogel 1 up to 7 days; a significant increase in cell number was found for each progressive time point. Cell status on TCPs is shown in Fig. S6 and S7 (Supp. Inf.) demonstrates cell morphology on hydrogel 1 from day 1 (Fig. S7a, Supp. Inf.) to day 7 (Fig. S7d, Supp. Inf.). L929 cells cultured on hydrogel 1 continued to spread and proliferate for the duration of the experiment, confirming the suitability of using hydrogel 1 as a material for wound diagnosis and to aid wound repair.

**Conclusions**

A redox-responsive, non-cytotoxic, 2% HA-based hydrogel was prepared through the facile crosslinking of HA with AED after 130 minutes at 37 °C. The hydrogels formed remained robust in aqueous solutions that lacked GSH, but lost structural integrity within four days when incubated in solution that contained GSH. This GSH-induced change altered the hydrogel properties both on microscopic and macroscopic levels. The distribution of RITC-dextran within hydrogels that were incubated in the presence of GSH revealed amplified fluorescent probe distribution in correspondence with increased GSH concentrations. Hydrogel crosslinker cleavage by GSH had a pronounced effect on hydrogel swelling; a 63% increase in hydrogel diameter was observed when the hydrogel was stored in 10 mM GSH simulated chronic wound fluid. Such hydrogel swelling was evident to the naked eye. The hydrogel offers a highly simplistic, label-free, method to monitor GSH concentration within a sample fluid, and may be used as a support for fibroblast growth and proliferation. Consequently, it is a promising candidate to be used for chronic wound diagnosis and to aid chronic wound repair.

**Conflicts of interest**

There are no conflicts to declare.

**Acknowledgements**

The authors would like to thank Algy Kazlauciunas and Ethan Perkins for experimental assistance and help with colorimetric assays. Sarah Myers and Matthew Percival are greatly acknowledged for training of Z.G. on cell culture and confocal laser microscopy.

# Supplementary Information

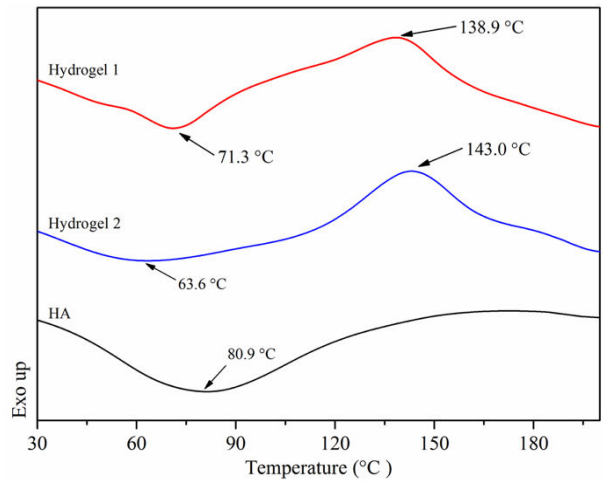

**Figure S1.** DSC thermograms corresponding to the crosslinked polymer used to form hydrogel **1** (top) and hydrogel **2** (middle), and linear HA.

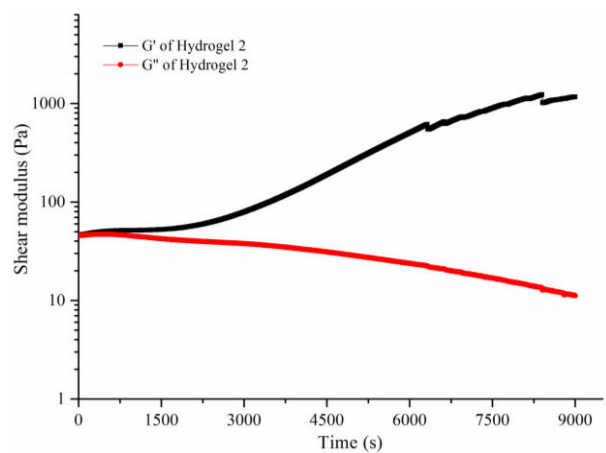

**Figure S2.** Time sweep rheological analysis during HA crosslinking to assess the gelation kinetics for hydrogel **2**.

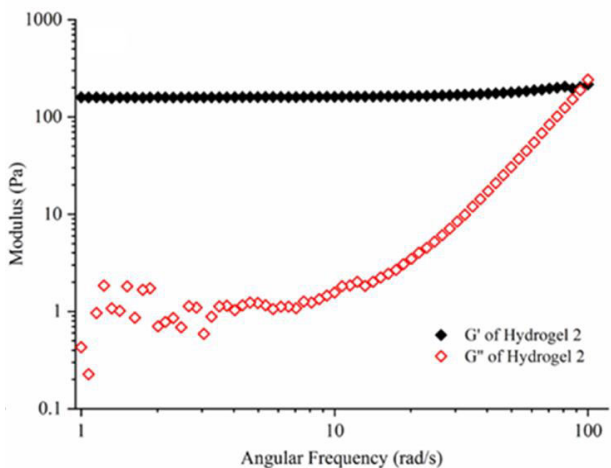

**Figure S3.** Frequency sweep rheological analysis of the storage ($G'$) and loss ($G''$) modulus in hydrogel **2**.

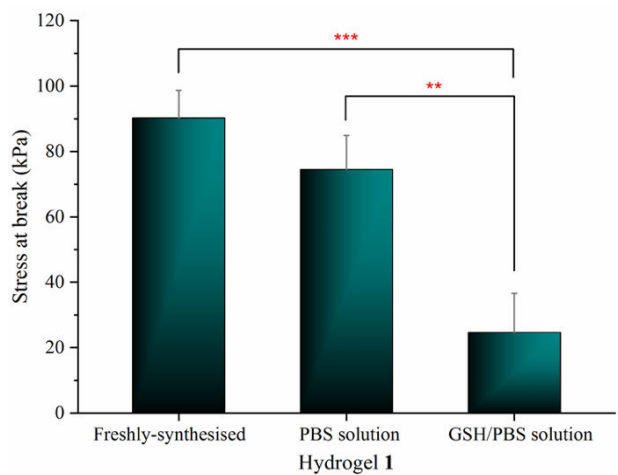

**Figure S4.** Stress at break measured for hydrogel **1** following synthesis and 48-hour incubation in either PBS or GSH (5 mM)-supplemented PBS buffer solution.

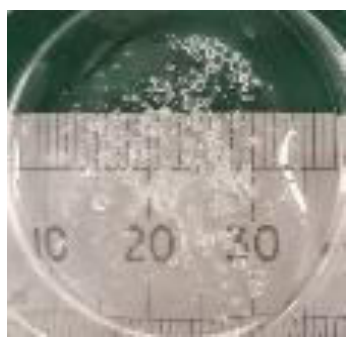

**Figure S5.** Dissolution of hydrogel **1** after 24-hour immersion in the simulated wound fluid containing 20 mM GSH. The material was too fragile to undergo any form of analysis.

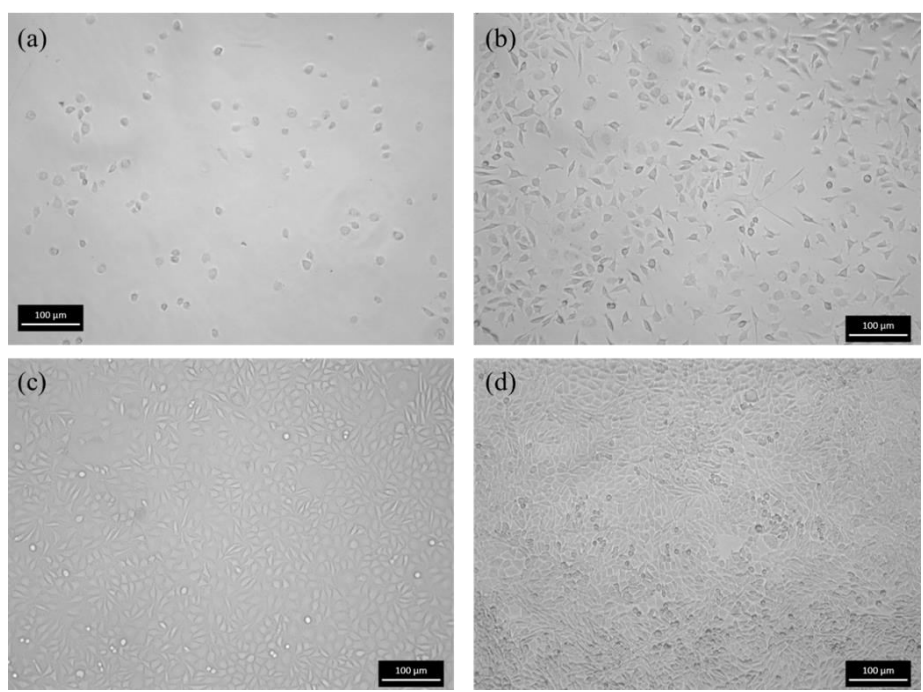

**Figure S6.** Optical microscopy images of L929 fibroblast cells cultured on TCP for 1 day (a), 3 days (b), 5 days (c) and 7 days (d).

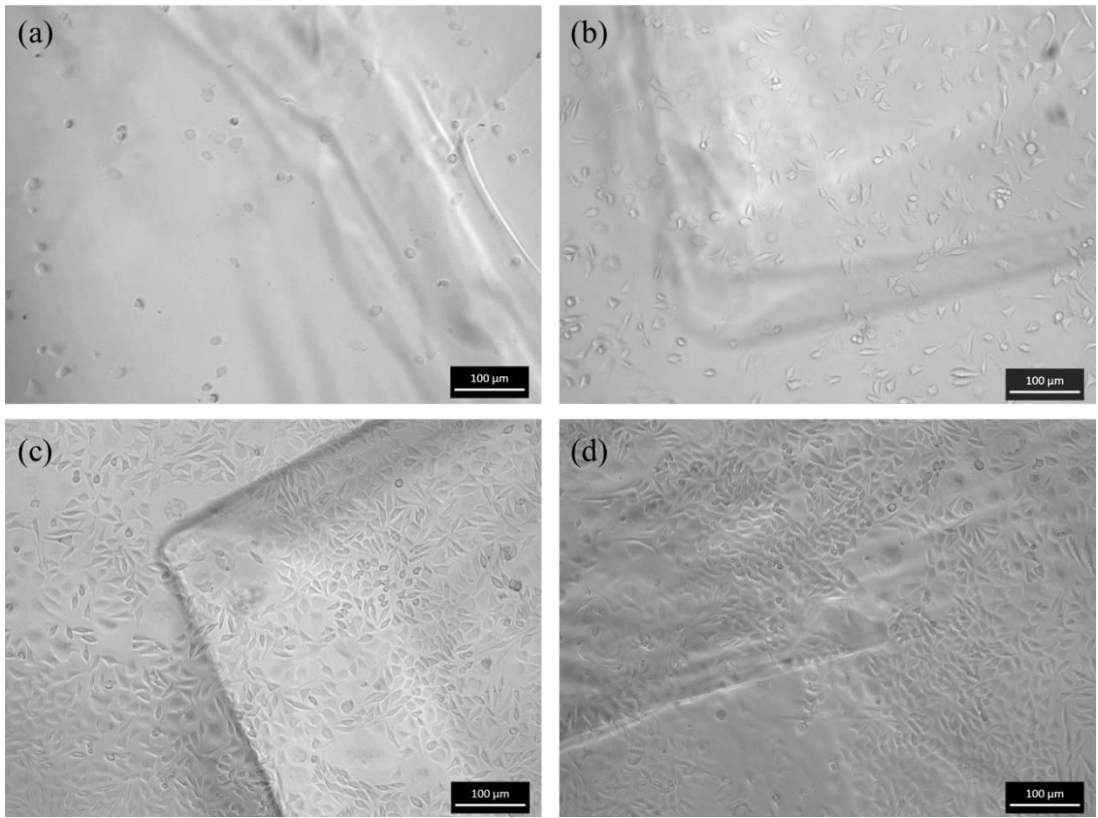

**Figure S7.** Optical microscopy images of L929 fibroblast cells cultured with hydrogel **1** for 1 day (a), 3 days (b), 5 days (c) and 7 days (d).

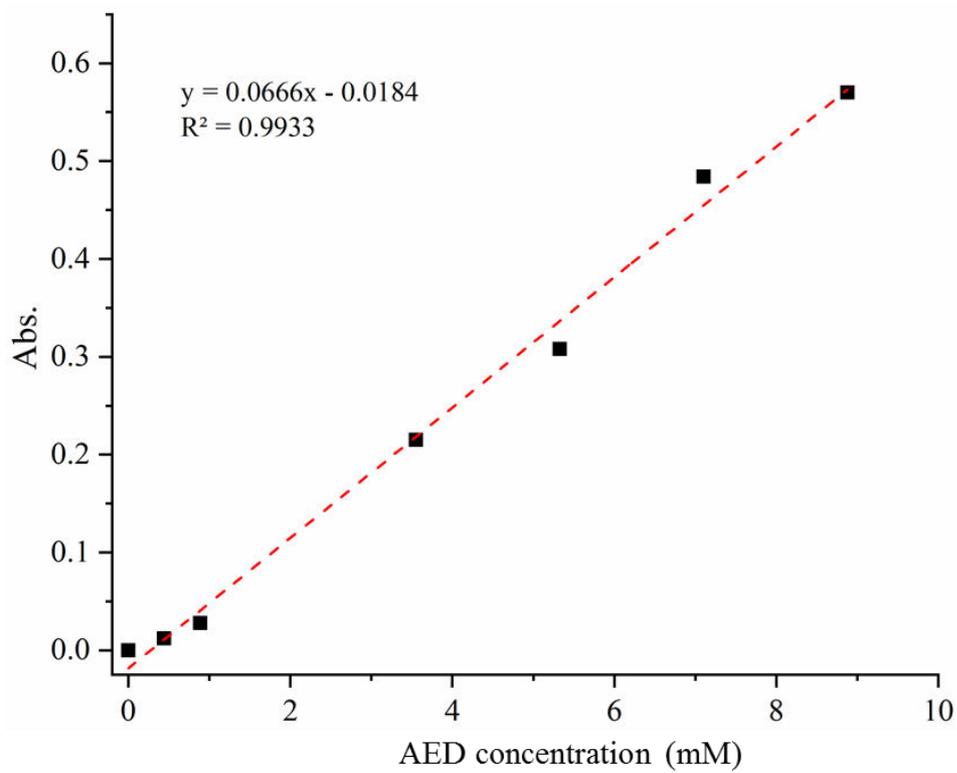

**Figure S8.** The calibration curve obtained for TNBS analysis.

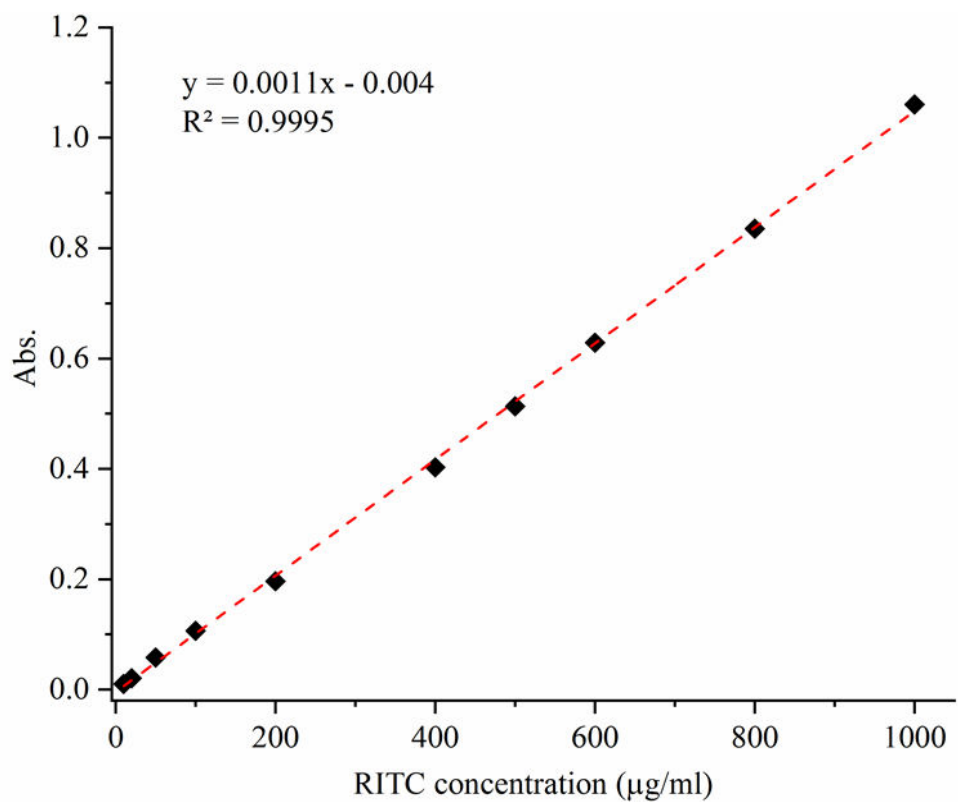

**Figure S9.** The calibration curve obtained for RITC analysis.

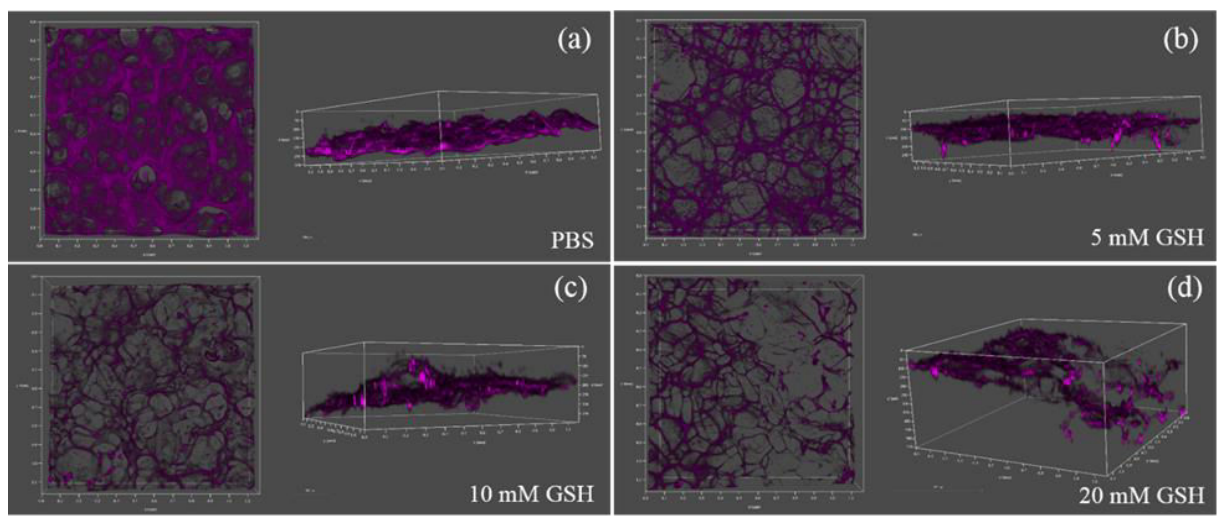

**Figure S10.** Confocal images hydrogel 1 loaded with RITC-dextran in (a) PBS solution, and PBS solution supplemented with (b) 5 mM GSH, (c) 10 mM GSH, and (d) 20 mM GSH. The stained polymer network is provided in the left-hand image of each box, and becomes increasingly disintegrated with increased GSH concentration. The depth to which RITC-dextran was able to penetrate the hydrogel was related fluorescence depth profile for each hydrogel is provided in the right-hand image of each box.